\documentclass[a4paper,fleqn,usenatbib,useAMS]{mnras}

\usepackage{graphicx}	
\usepackage{amsmath}	
\usepackage{amssymb}	

\usepackage[T1]{fontenc}
\usepackage{ae,aecompl}

\usepackage{newtxtext,newtxmath}

\title[Spin-down rate of PSR J1023+0038 with Aqueye+]{Spin-down rate of the transitional millisecond pulsar PSR J1023+0038 in the optical band with Aqueye+} 

\author[A. Burtovoi, L. Zampieri, M. Fiori et al.]{
Aleksandr Burtovoi$^{1,2}$\thanks{E-mail: aleksandr.burtovoi90@gmail.com},
Luca Zampieri$^{2}$\thanks{E-mail: luca.zampieri@inaf.it},
Michele Fiori$^{3,2}$,
Giampiero Naletto$^{3,2}$,
\newauthor
Alessia Spolon$^{3,2}$,
Cesare Barbieri$^{3,2}$,
Alessandro Papitto$^{4}$
and Filippo Ambrosino$^{5}$
\\
$^{1}$Centre of Studies and Activities for Space (CISAS) ``G. Colombo'', University of Padova, Via Venezia 15, 35131 Padova, Italy\\
$^{2}$INAF - Osservatorio Astronomico di Padova, Vicolo dell'Osservatorio 5, 35122, Padova, Italy\\
$^{3}$Department of Physics and Astronomy, University of Padova, Via F. Marzolo 8, 35131, Padova, Italy\\
$^{4}$INAF - Osservatorio Astronomico di Roma, Via Frascati 33, 00076, Monteporzio Catone (RM), Italy\\
$^{5}$INAF - Istituto di Astrofisica e Planetologia Spaziali, Via del Fosso del Cavaliere 100, 00133 Rome, Italy 
}

\date{Accepted XXX. Received YYY; in original form ZZZ}

\pubyear{2020}

\begin{document}
\label{firstpage}
\pagerange{\pageref{firstpage}--\pageref{lastpage}}
\maketitle

\begin{abstract}
We present a timing analysis of the transitional millisecond pulsar PSR J1023+0038 using observations taken between January 2018 and January 2020 with the high time resolution photon counter Aqueye+ mounted at the 1.82 m Copernicus telescope in Asiago. We report the first measurement of the timing solution and the frequency derivative of PSR J1023+0038 based entirely on optical data. The spin-down rate of the pulsar is $(-2.53 \pm 0.04) \times 10^{-15}$ Hz$^2$, which is $\sim$20\% slower than that measured from the X-ray observations taken in 2013-2016 and $\sim$5\% faster than that measured in the radio band during the rotation-powered state.
\end{abstract}




\section{Introduction}\label{sec:1}
Millisecond pulsars (MSPs) are believed to be rather old and fast rotating neutron stars formed in binary systems which then spun-up to millisecond periods during long-term accretion from a companion star (see e.g. \citealt{Alpar1982, Radhakrishnan1982}). MSPs are usually observed either as accreting X-ray pulsars or as rotation-powered radio pulsars with no ongoing accretion. Recently, it was discovered that a few members of the MSP population, called transitional millisecond pulsars (tMSPs), show an amazing behavior. They switch between an accretion-powered and a rotation-powered regime. At present, we know three sources that behave in this way: PSR J1023+0038 \citep{Archibald2009}, PSR J1227$-$4853 \citep{deMartino2010} and PSR J1824$-$2452 \citep{Papitto2013}.

PSR J1023+0038 is a transitional millisecond pulsar currently in a Low Mass X-ray Binary (LMXB) state. So far, it is the only tMSP which has shown detectable pulsed emission in the optical band. Optical pulsations were discovered with SIFAP at the Telescopio Nazionale Galileo \citep{Ambrosino2017} and soon confirmed with Aqueye+ at the Copernicus telescope in Asiago \citep{Zampieri2019}, and also with the panoramic photometer-polarimeter mounted at BTA in Nizhniy Arkhyz \citep{Karpov2019}. \citet{Papitto2019} studied in detail the properties of the optical and X-ray pulses performing the first high time resolution multi-wavelength observational campaign of this source.

The exact mechanism of these pulsations is still under discussion. Different models have been proposed to explain it (see e.g. \citealt{Campana2019, Papitto2019, Veledina2019} and references therein for a detailed review of the theoretical models). In order to improve our understanding on the properties of optical pulsations and the underling emission mechanism, it is important to increase the number, accuracy and baseline of the available optical measurements.

In our previous work \citep{Zampieri2019} we demonstrated the capability of the fast photon counter Aqueye+ in detecting significant optical pulsations from PSR J1023+0038 and in deriving an independent optical timing solution over a baseline of a few days. In this work we extend this study, performing a detailed timing analysis of five runs of Aqueye+ taken in 2018-2020. Our main goal is to find an accurate timing solution and to measure the frequency derivative of PSR J1023+0038.

In Section \ref{sec:2} we present the details on the Aqueye+ observations and the data reduction. The procedure used to perform the correction for the orbital motion, along with the results on the timing analysis are reported in Section \ref{sec:3}. The discussion and conclusions follow in Section \ref{sec:4}.

\section{Observations and Data Reduction}\label{sec:2}
We carried out five observing runs of PSR J1023+0038 with Aqueye+\footnote{\url{https://web.oapd.inaf.it/zampieri/aqueye-iqueye/index.html}} fast photon counter \citep{Barbieri2009, Naletto2013, Zampieri2015} mounted at the 1.8 m Copernicus telescope (Asiago, Italy) during three years since January 2018 (see the summary of the observations in Table \ref{tab:log_dtasc}).

\begin{table}
\scriptsize
\centering
\caption{Summary of the Aqueye+ observations of PSR J1023+0038 taken at the 1.8 m Copernicus telescope in Asiago during five runs in 2018, 2019 and 2020. Start times refer to the Solar system barycenter. The total (non-continuous) on-source observing time $T_{\rm obs}$ for each night is listed in the third column. Correction $\Delta T_{\rm asc}$ and the time of the ascending node passage $T_{\rm asc}$ calculated from the epoch folding search are shown in the last two columns.}
\label{tab:log_dtasc}
\begin{tabular}{l c r c l}
\hline
\hline
Obs. night & Start time & $T_{\rm obs}$ & $\Delta T_{\rm asc}$$^a$ & \multicolumn{1}{c}{$T_{\rm asc}$} \\
& (MJD)  & (ks) & (s) & \multicolumn{1}{c}{(MJD)} \\
\hline
\multicolumn{5}{l}{\textit{Run 1}} \\
Jan 22, 2018	& 58140.0140070719	& 12.6	& $10.87 \pm 0.26$ & 58139.893489(3) \\
Jan 23, 2018	& 58141.0322954264	& 10.8	& $11.18 \pm 0.23$ & 58140.883974(3) \\
Jan 24, 2018	& 58142.0703056106	& 10.8	& $11.56 \pm 0.18$ & 58141.874460(2) \\
Jan 25, 2018	& 58143.0399887431	& 9.9	& $11.55 \pm 0.07$ & 58142.8649416(8) \\	
\noalign{\vskip 0.25mm} 
\multicolumn{5}{l}{\textit{Run 2}} \\
Dec 11, 2018	& 58463.0679998575	& 9		& $22.92 \pm 0.16$ & 58462.988719(2) \\
Dec 12, 2018	& 58464.0446229072	& 13.5	& $22.93 \pm 0.08$ & 58463.9792008(9) \\
Dec 13, 2018	& 58465.0343255807	& 12		& $23.17 \pm 0.20$ & 58464.969685(2) \\
Dec 14, 2018	& 58466.0467962778	& 11.7	& $22.63 \pm 0.17$  & 58465.960160(2) \\
Dec 15, 2018	& 58467.0313471542	& 13.2	& $22.88 \pm 0.15$ & 58466.950645(2) \\
\noalign{\vskip 0.25mm} 
\multicolumn{5}{l}{\textit{Run 3}} \\
Feb 4-5, 2019	& 58518.9578030587	& 13.5	& $24.13 \pm 0.29$ & 58518.851894(3) \\
Feb 5-6, 2019	& 58519.8790642319	& 19.8	& $24.70 \pm 0.07$ & 58519.8423822(8) \\
Feb 6-7, 2019	& 58520.8774679789	& 18		& $24.45 \pm 0.06$ & 58520.8328609(7) \\
\noalign{\vskip 0.25mm} 
\multicolumn{5}{l}{\textit{Run 4}} \\
Nov 26, 2019	& 58813.0771640569	& 10.8	& $33.03 \pm 0.07$ & 58813.0250256(8) \\
\noalign{\vskip 0.25mm} 
\multicolumn{5}{l}{\textit{Run 5}} \\
Jan 27, 2020	& 58875.0191847258	& 10.8	& $33.50 \pm 0.10$ & 58874.831081(1) \\
Jan 28-29, 2020	& 58876.9346872752	& 18.3	& $33.59 \pm 0.16$ & 58876.812046(2) \\
Jan 29-30, 2020	& 58877.9332196822	& 12.6	& $33.37 \pm 0.07$ & 58877.8025247(8) \\
Jan 31, 2020	& 58878.9690555457	& 9		& $33.29 \pm 0.17$ & 58878.793005(2) \\
\hline
\end{tabular}
\begin{minipage}{8.1 cm}
$^a$ The uncertainty on $\Delta T_{\rm asc}$ is the square root of the diagonal term of the covariance matrix of the fit which corresponds to the centroid of the Gaussian function. 
\end{minipage}
\end{table} 

The data have been reduced with the \texttt{QUEST} software (v. 1.1.5, see \citealt{Zampieri2015}). Arrival time of each photon was referred to the solar system barycenter using the \texttt{TEMPO2} package in TDB units \citep{Edwards2006, Hobbs2006}, using the JPL ephemerides DE405. The position of PSR J1023+0038 was taken from \citet{Jaodand2016}: $\alpha=10^{\rm h}23^{\rm m}47.687198^{\rm s}$, $\delta=+00^\circ38^\prime40.84551^{\prime\prime}$ at epoch MJD 54995.

We modified the barycentered time series correcting the photon arrival times for the pulsar orbital motion. The orbital parameters are taken from \citet{Jaodand2016}: $a/c=0.343356$ s, $P_{\rm orb} = 0.1980963155$ d. The reference date for the ascending node passage is MJD 57449.7258 \citep{Ambrosino2017}. We assume no variation of the orbital period of the system with time. All variations of the orbital parameters are accounted for changing the time of the ascending node passage (see \citealt{Jaodand2016}).

\section{Analysis and Results}\label{sec:3}
\subsection{Search for the epoch of the ascending node passage}\label{sec:3.2}
During the accretion state of PSR J1023+0038 the epoch of the ascending node passage $T_{\rm asc}$ is known to show significant variations with time of the order of several seconds (see e.g. \citealt{Jaodand2016, Papitto2019}). To account for this, we performed an accurate epoch folding search for $T_{\rm asc}$ for each observing night. We folded the baricentred time series corrected for the binary motion assuming different values of $T_{\rm asc}$, with the aim to find the one which gives the pulse profile with the highest $\chi^2$. The folding period for each night was calculated extrapolating the X-ray ephemerides from \citet{Jaodand2016}. Combing the data accumulated during a full observing night (i.e. several hours of observation) allows us to detected the pulse profile with a significance of >3.2-$\sigma$ ($\chi^2 \gtrsim37$ for 15 degrees of freedom (d.o.f.), 16 phase bins).

In order to determine $T_{\rm asc}$, we first performed a preliminary search within an interval of $\pm$2 min around the expected value in steps of 2 s. Then, a finer search was carried out within $\pm$10 s around previously estimated value, using steps of 0.5 s. The final value of the correction $\Delta T_{\rm asc}$ to the time of the ascending node was determined by fitting the peaks of the $\chi^2$ distribution with a Gaussian function (see Table \ref{tab:log_dtasc}). The uncertainty on $\Delta T_{\rm asc}$ is calculated as the square root of the diagonal term of the covariance matrix of the fit which corresponds to the centroid of the Gaussian function.

Since $T_{\rm asc}$ does not vary significantly on a time scale of several days, for all the observations of the same run we adopted the same value of $\Delta T_{\rm asc}$, determined for the observing night with the highest $\chi^2$. For the January 2018 run $\Delta T_{\rm asc}$ is equal to 11.55 s (as inferred from the January 25 data), for the December 2018 run it is equal to 22.93 s (as inferred from the December 12 data), for the February 2019 run $\Delta T_{\rm asc}=24.45$ s (as inferred from the February 6-7 data), for the November 2019 run $\Delta T_{\rm asc}=33.03$ s (as inferred from the November 26 data) and for the January 2020 run $\Delta T_{\rm asc}=33.37$ s (as inferred from the January 29-30 data).

\begin{figure}
	\centering
	\includegraphics[width=0.5\textwidth]{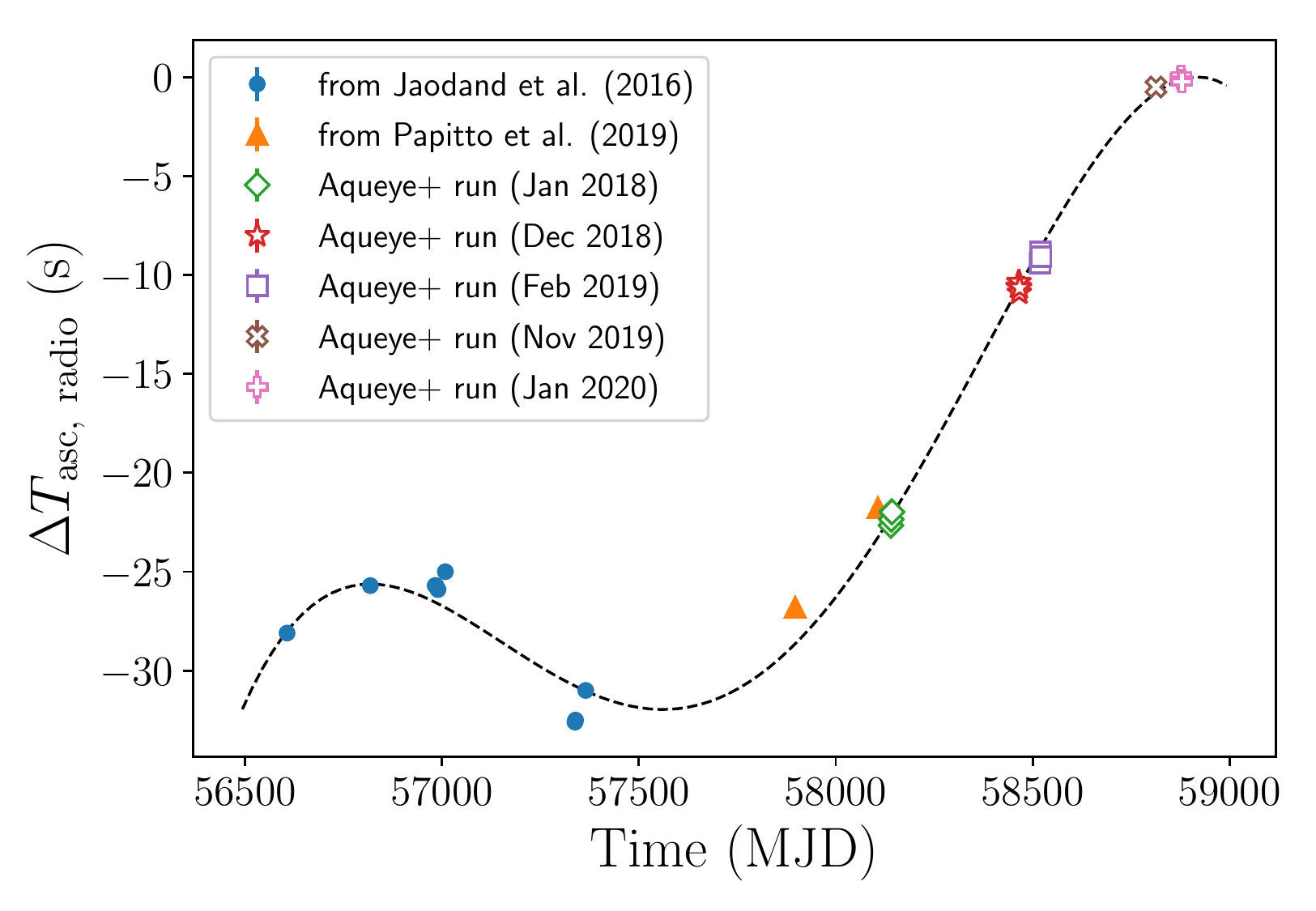} \\
	\caption{Evolution of the time of the ascending node passage $T_{\rm asc}$. $\Delta T_{\rm asc,\, radio}$ is defined as a difference between the actual value of $T_{\rm asc}$ and that calculated using radio timing solution \citep{Archibald2013arx, Jaodand2016}. The dashed line shows a fourth-order polynomial function fit to the $\Delta T_{\rm asc,\, radio}$ measurements.
}
	\label{fig:dtasc_radio}
\end{figure}

We calculated also the difference $\Delta T_{\rm asc,\, radio}$ between our value of the time of the ascending node passage and that calculated using the radio timing solution \citep{Archibald2013arx, Jaodand2016} at the epoch of the Aqueye+ observations. The long-term evolution of $\Delta T_{\rm asc,\, radio}$ shown in Fig. \ref{fig:dtasc_radio} can be reasonably well described by a fourth-order polynomial function. Similar figures have been published earlier (see e.g. \citealt{Jaodand2016, Papitto2019}). However, the latest Aqueye+ measurements show an increasing trend of $\Delta T_{\rm asc,\, radio}$ with time since 2017 May (MJD 57896). 
Although the overall evolution of $T_{\rm asc}$ is still consistent with a random (red-noise) process, the continuous steady increase that we detected between 2018 and 2020, if confirmed with future observations, could in fact indicate a systematic underestimate of the orbital period of the system.

\subsection{Analysis of the individual Aqueye+ runs}\label{sec:3.4}
For each run we performed a phase fitting of the Aqueye+ data corrected for the binary motion using the values of $T_{\rm asc}$ determined as explained in Section \ref{sec:3.2}. We folded separately each night of observations acquired during a single run (Table \ref{tab:log_dtasc}) using 16 phase bins. A different reference period $P_{\rm init}$ was adopted for folding the data of each run. The value of $P_{\rm init}$ was calculated extrapolating the X-ray ephemerides from \citet{Jaodand2016} at the beginning of the run. 

\begin{figure}
	\centering
	\includegraphics[width=0.49\textwidth]{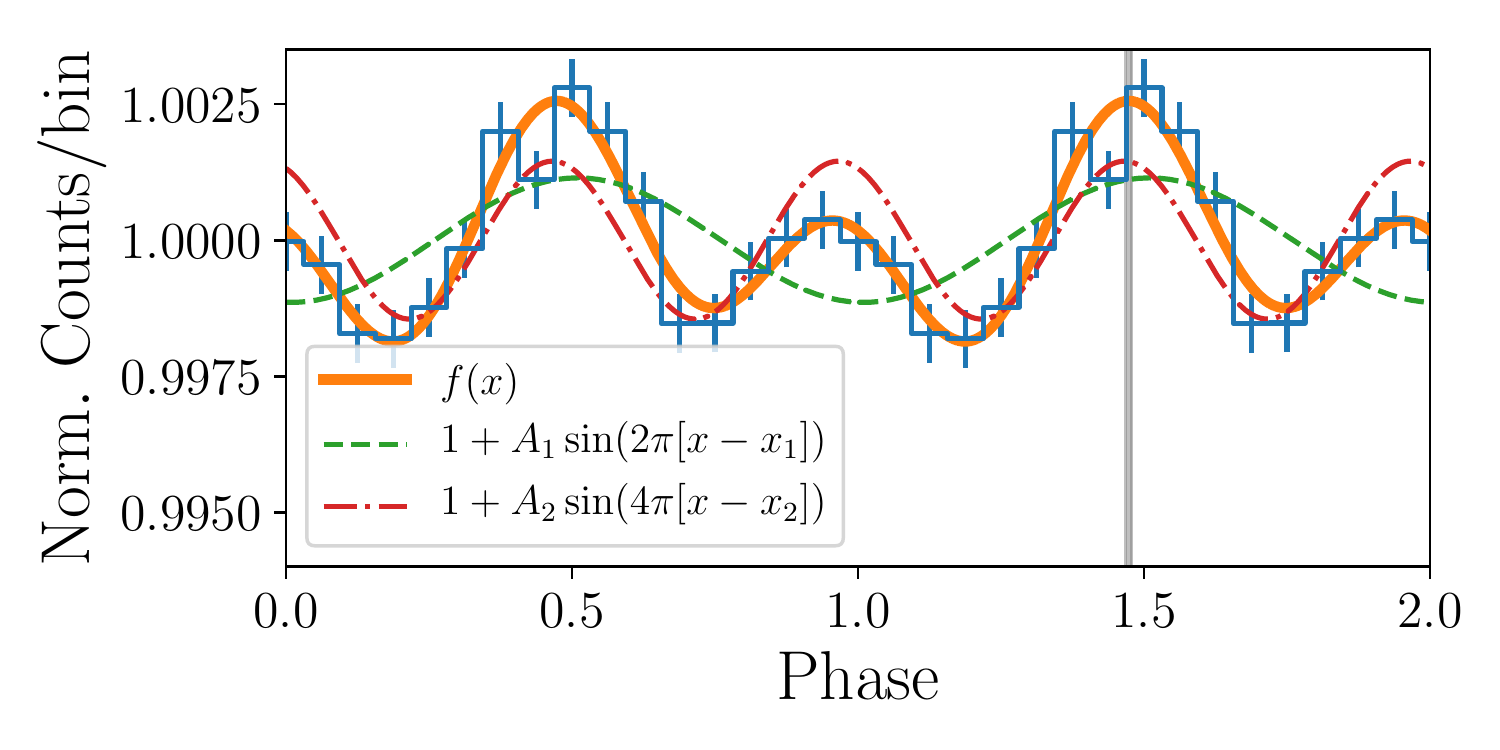} \\
	\caption{Pulse profile of PSR J1023+0038 obtained folding one night of observations (January 25, 2018) using the reference period $P_{\rm init}=1.687987440$ ms and 16 phase bins. Two rotational phases are shown. The solid line shows the fitting function $f(x)$ (equation \ref{eq:2sin_fun}). The dashed and dot-dashed lines represent the two harmonically-related sinusoids. The gray area, which is rather narrow, marks the position of the main peak, including its estimated uncertainty.}
	\label{fig:pp_25Jan2018}
\end{figure}

The folded pulse profiles were then fit with an analytical template, which reproduces the shape of the pulse profile of PSR J1023+0038. Following \citet{Ambrosino2017, Zampieri2019}, we adopted the sum of two harmonically-related sinusoids plus a constant (see , e.g., Fig. \ref{fig:pp_25Jan2018}):
\begin{equation}
	f(x) = K \, \{ 1 + A_1 \sin(2\pi[x-x_1]) + A_2 \sin(4\pi[x-x_2]) \}\,.
	\label{eq:2sin_fun}
\end{equation}

In order to track the phase of the pulsar, we determined the position of the most prominent peak of each pulse profile. We calculated the local maxima of the best fitting function $f$ computing the zeros of the derivative $f' \equiv df/dx$. Uncertainties of the peak position have been estimated from the difference of the zeros of the functions $f' \pm \delta f'$ and $f'$, where $\delta f'$ was obtained propagating the errors of the fitted parameters $p_i$ on $f'$:
\begin{equation}
        \delta f' = \sqrt{\sum_i \sum_j \left(\frac{\partial f'}{\partial p_i}\right) \left(\frac{\partial f'}{\partial p_j}\right) \sigma_{ij}}.
        \label{eq:df_deriv}
\end{equation}
$\sigma_{ij}$ is the covariance matrix of the fit, whereas the parameters $p_i$ correspond to $x_1$, $x_2$, $A_1$, $A_2$ and $K$. The typical uncertainty on the position of the peak determined in this way is $\sim$0.007, or $\sim$12 $\mu$s.

The peak positions obtained for each night of the same run were then fit with a phase function of the form:
$\psi(t) = \phi(t) - (t-t_0)/P_{\rm init}$ \citep{Zampieri2014}. 
In this equation $\phi(t)$, the actual rotational phase of the pulsar, is modelled for each run as a first order polynomial: $\phi(t) = \phi_0 + \nu_0(t-t_0)$, where $\phi_0$ and $\nu_0$ are the phase and frequency of the pulsar at the reference time $t_0$. The resulting timing solutions, with the final fitted values of $\phi_0$ and $\nu_0$ for all Aqueye+ runs (except the November 2019 run) are listed in Table \ref{tab:tss}. Since only one night of good data has been acquired in November 2019, we were not able to obtain a local timing solution for this run.

The parameters $\phi_0$ and $\nu_0$ of the timing solution obtained for the January 2018 run are consistent with those reported in \citet{Zampieri2019}. The present fit has slightly smaller uncertainties in $\phi_0$ and $\nu_0$ than those reported in \citet{Zampieri2019}, because we adopt a linear in place of a quadratic spin-down law. In addition, since there is evidence that the ratio of the amplitudes of two sinusoids ($A_2/A_1$) varies from night to night, we left it free in the present fit, while it was kept fixed in \citet{Zampieri2019}.

We note that the reduced $\chi^2$ calculated for the linear fit of the December 2018 and January 2020 runs is rather high (12.4 and 6.2, respectively). The significant scatter of the phase measurements is likely caused by the intrinsic timing noise of the pulsar and/or by uncertainties in the values of the orbital parameters (see Section \ref{sec:4}).
To account for this, as better estimate of the actual uncertainty of the phase measurement we considered the dispersion (standard deviation) of the phase measurements around the best fitting linear model, which is equal to 0.02 for the December 2018 and 0.01 for the January 2020 run. Refitting the pulsar phases, we obtained a value of the reduced $\chi^2$ close to 1. The best fitting parameters are consistent with those reported in Table \ref{tab:tss}.

\begin{table*}
\centering
\caption{Timing solutions of PSR J1023+0038 for the four separate Aqueye+ observing runs.}
\label{tab:tss}
\begin{tabular}{l l l l l}
\hline
\hline
 					& Jan 2018 run $^a$ 					& Dec 2018 run $^b$					& Feb 2019 run $^c$						& Jan 2020 run $^d$	\\
\hline
$t_0$ (MJD)			& 58140 									& 58463									& 58518									& 58875									\\
$\phi_0$			& $0.108 \pm 0.002$							& $0.26 \pm 0.02$						& $0.799 \pm 0.008$						& $0.242 \pm 0.017$						\\
$\nu_0$ (Hz)		& $592.42146753 \pm 1 \times 10^{-8}$		& $592.42146760 \pm 11 \times 10^{-8}$	& $592.42146746 \pm 4 \times 10^{-8}$	& $592.42146750 \pm 7 \times 10^{-8}$	\\
$P_0$ (ms)			& $1.68798744612 \pm 0.3 \times 10^{-10}$	& $1.6879874459 \pm 3 \times 10^{-10}$	& $1.6879874463 \pm 1 \times 10^{-10}$	& $1.6879874462 \pm 2 \times 10^{-10}$	\\
$\chi^2$/d.o.f. $^e$	& 0.3/(4-2)									& 37.1/(5-2)								& 0.8/(3-2)								& 12.3/(4-2)								\\
timing noise $^f$		& 0.005										& 0.045									& 0.005									& 0.023									\\
\hline
\end{tabular}
\begin{minipage}{17 cm}
$^a$ $P_{\rm init}=1.687987440$ ms. 
$^b$ $P_{\rm init}=1.68798744645$ ms. 
$^c$ $P_{\rm init}=1.68798744649$ ms. 
$^d$ $P_{\rm init}=1.68798744675$ ms. 
$^e$ $\chi^2$ value of the linear phase fit. 
$^f$ Timing noise is calculated as the sum of the residuals of the linear fit (in quadrature).
\end{minipage}
\end{table*}

We also tried to fit the measured phases of each single run with a second-order polynomial function, but we have not been able to determine the frequency derivative $\dot\nu$. Only an upper limit was derived ($|\dot\nu| \lesssim 1 \times 10^{-12}$ Hz$^2$).

\subsection{Three-year timing solution from Aqueye+ data}\label{sec:3.5}
We combined the data from all five observing runs of Aqueye+ and obtained a timing solution valid from January 2018 through January 2020. We measured the phases of the main peak as described in Section \ref{sec:3.4}, folding all the data with common reference period $P_{\rm init} = 1.68798744634$ ms and reference time $t_0=58518$ MJD.
We conservatively adopted the standard deviation of the phases of the December 2018 run (0.02, see Section \ref{sec:3.4}) as the errors on all phase measurements. 

\begin{figure}
\centering
	\includegraphics[width=0.5\textwidth]{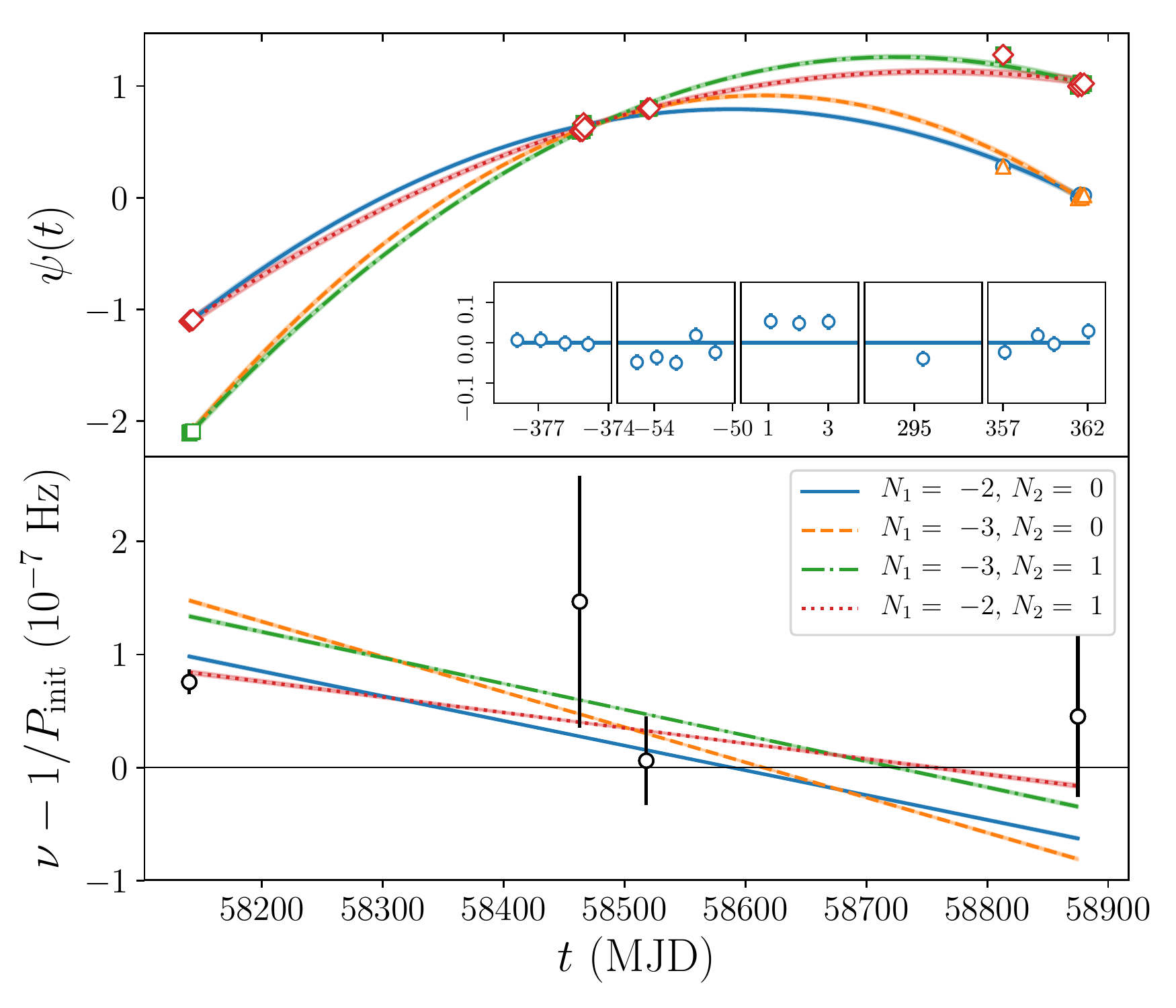}
	\caption{\textit{Top panel}: Phases $\psi(t)$ of PSR J1023+0038 measured over five Aqueye+ runs with respect to a uniform rotation with period $P_{\rm init}$, fit with a second-order polynomial function. The integer number of cycles [$N_1$, $N_2$] is assumed to be equal to [$-$2, 0] (blue solid line and circles), [$-$3, 0] (orange dashed line and triangles), [$-$3, 1] (green dot-dashed line and squares) and [$-$2, 1] (red dotted line and diamonds). The insets show residuals of the fit for [$N_1$, $N_2$] equal to [$-$2, 0] with the x-axis in days starting from MJD 58518. \textit{Bottom panel:} The spin evolution $\nu(t)$ derived from the final timing solution for different values of $N_1$ and $N_2$. The butterfly plots, which are rather narrow, are calculated propagating the 1-$\sigma$ uncertainties. The values of the frequency obtained from the individual timing solutions of each run are plotted as black open circles (see Table \ref{tab:tss}).}
	\label{fig:tot_tss}
\end{figure}

In order to obtain the three-year timing solution, we fit the measured phases using a second-order polynomial function: $\psi(t) = \phi_0 + (\nu_0 - 1/P_{\rm init})\,(t-t_0) + (\dot\nu_0/2)\,(t-t_0)^2$. 
Since the integer numbers of cycles between runs separated by more than $\sim$3 months are uncertain (run 1-2 and 3-4 in Table \ref{tab:log_dtasc}), we determined them minimizing the $\chi^2$ of the phase fit. 
The four timing solutions with the lowest values of the $\chi^2$ are shown in Fig. \ref{fig:tot_tss} and reported in Table \ref{tab:chi2_list}, where $N_1 = {\rm int}[\psi_{\rm MJD~58518}-\psi_{\rm MJD~58140}]$ and $N_2 = {\rm int}[\psi_{\rm MJD~58813}-\psi_{\rm MJD~58518}]$. Although the residuals of these timing solutions are rather high ($\sigma_{\rm rej}>3$), they provide a reasonable representation of the average evolution of the pulsar phase. Fitting the phases with other combinations of $N_1$ and $N_2$ resulted in even higher residuals ($\chi^2>100$ with $\sigma_{\rm rej} > 8$) and, therefore, we did not consider them further in our analysis. 

\begin{table}
\centering
\caption{The values of $\chi^2$ of the phase fit for different values of $N_1$ and $N_2$. The last two columns list the sum of the squares of the differences between $\nu(t)$ and the individual measurements of the rotational frequency obtained in the X-ray and optical bands divided by the corresponding squared uncertainties ($\chi^2_{\nu,\rm (x+opt)}$, see text for details), and the value of the frequency derivative inferred from the fit ($\dot \nu_0$). The rejection probability calculated from the chi-square distribution is reported in terms of Gaussian sigma ($\sigma_{\rm rej}$). The number of degrees of freedom is equal to 14 (=17-3) for $\chi^2$ and 8 for $\chi^2_{\nu,\rm (x+opt)}$.}
\label{tab:chi2_list}
\begin{tabular}{c c c c c}
\hline\hline
$N_1$	& $N_2$	& $\chi^2$ ($\sigma_{\rm rej}$) 	& $\chi^2_{\nu,\rm (x+opt)}$ ($\sigma_{\rm rej}$)	& $\dot \nu_0$ [$10^{-15}$ Hz$^2$] \\
\hline
$-$2	& 0	& 45.16 (4.1-$\sigma$)	& 11.93 (1.4-$\sigma$)	& $-2.53 \pm 0.04$ \\
$-$3	& 0	& 47.45 (4.3-$\sigma$)	& 98.30 ($>$8-$\sigma$)	& $-3.60 \pm 0.04$ \\
$-$3	& 1	& 59.98 (5.3-$\sigma$)	& 37.55 (4.4-$\sigma$)	& $-2.65 \pm 0.04$ \\
$-$2	& 1	& 90.41 (7.3-$\sigma$)	& 44.41 (5.0-$\sigma$)	& $-1.58 \pm 0.05$ \\
\hline\hline
\end{tabular}
\end{table}

\begin{figure}
\centering
	\includegraphics[width=0.5\textwidth]{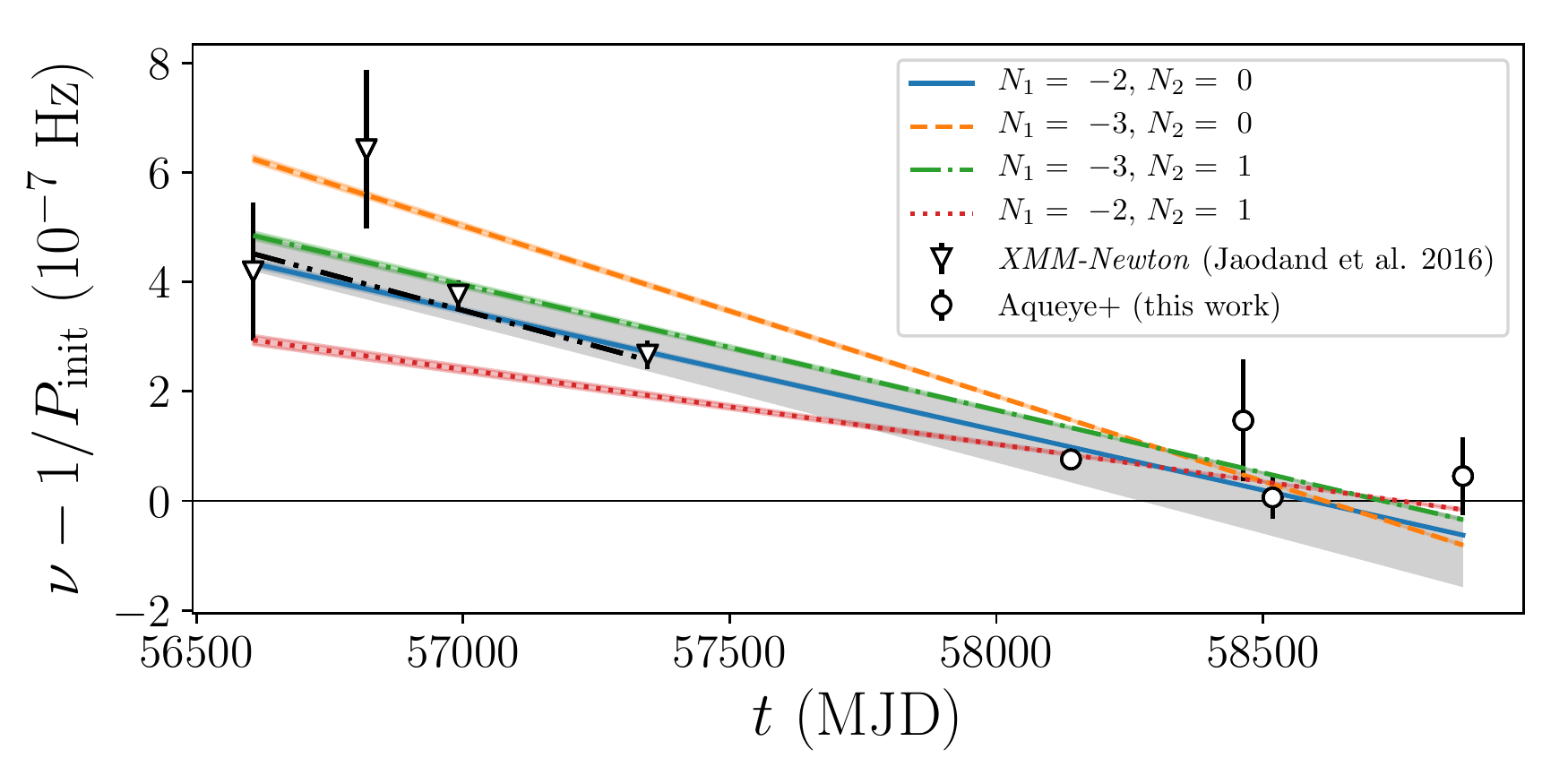}
	\caption{Same as Fig. \ref{fig:tot_tss} (bottom panel) overlaid with the X-ray measurements (black open triangles) of the frequency adopted from \citet{Jaodand2016}. The black double-dot dashed line shows the spin evolution $\nu(t)$ derived from the best fitting X-ray timing solution. The gray area corresponds to the butterfly plot of the joint linear fit of the X-ray and optical frequency measurements.}
	\label{fig:freq_aq_j16}
\end{figure}

In order to place an additional constraint on the frequency derivative and the timing solution, we added the past \textit{XMM-Netwon} measurements of the rotational frequency \citep{Jaodand2016} to the Aqueye+ measurements. In this way, we tested whether the long-term evolution of the spin frequency during the interval of time covering the X-rays and optical observations ($\sim$6 years) is consistent with the trend inferred from the timing solutions reported in Table \ref{tab:chi2_list}. For each timing solution we calculated the rotational frequency $\nu(t)$ and compared it with the values reported in \citet{Jaodand2016} and those measured for each individual optical run (see Fig. \ref{fig:freq_aq_j16}).

We calculated the sum $\chi^2_{\nu,\rm (x+opt)}$ of the square differences between $\nu(t)$ and the individual measurements of the rotational frequency divided by the corresponding squared uncertainties. As shown in Table \ref{tab:chi2_list}, the timing solution calculated for $N_1=-2$ and $N_2=0$ shows the best agreement between the long-term and ``local'' values of the spin frequencies. 
In the assumption that the frequency noise is not significant, this solution is then favored. The resulting parameters of this timing solution are listed in Table \ref{tab:5runs_tss}. 
The values of the frequency and its derivative are consistent with those obtained performing a joint linear fit of the X-ray and optical frequency measurements (see the gray area in Fig. \ref{fig:freq_aq_j16}): $\nu_{\rm fit}(t_1) = 592.42146790 \pm 3 \times 10^{-8}~{\rm Hz}$ and $\dot\nu_{\rm fit}(t_1) = (-2.8 \pm 0.3) \times 10^{-15}~{\rm Hz}^2$, where $t_1=56606.6$ MJD.

\begin{table}
\centering
\caption{Quasi-coherent three-year timing solution of PSR J1023+0038 for five observing runs with Aqueye+, calculated for $N_1=-2$ and $N_2=0$.}
\label{tab:5runs_tss}
\begin{tabular}{l l}
\hline\hline
\multicolumn{2}{c}{All Aqueye+ observations in Jan 2018 -- Jan 2020}	\\
\hline
$\phi_0$				& $0.745 \pm 0.013$							\\
$\nu_0$ (Hz)			& $592.4214674668 \pm 4 \times 10^{-10}$	\\
$\dot\nu_0$ (Hz$^2$)	& $(-2.53 \pm 0.04) \times 10^{-15}$			\\
$P_0$ (ms)				& $1.6879874462956 \pm 1.1 \times 10^{-12}$\\
$\chi^2$/d.o.f. $^a$		& 45.16/(17-3)								\\
$\chi^2_{\nu,\rm (x+opt)}$/d.o.f. $^b$& 11.93/8						\\
\hline
\end{tabular}
\begin{minipage}{8 cm}
$P_{\rm init} = 1.68798744634$ ms and $t_0 = 58518$ MJD. The timing noise, calculated as the sum of the residuals from the fitted model in quadrature, is equal to $\sim$0.14. $^a$ $\chi^2$ value of the parabola fit of the measured phases. 
$^b$ The $\chi^2_{\nu,\rm (x+opt)}$ value is calculated as the sum of the square differences between $\nu(t)$ and the individual measurements of the rotational frequency obtained in the X-ray and optical bands (see text for details) divided by the corresponding squared uncertainties.
\end{minipage}
\end{table}

We tried to perform a fit of the phase measurements using a third-order polynomial function (introducing $\ddot\nu$). Although we obtained several timing solutions with statistically satisfactory phase fits, none of them is able to reproduce well the long-term evolution of the spin frequency.

In an attempt to characterize the noise of the phase measurements around the best timing solution ([$-$2, 0]), 
we fit the corresponding phase residuals with a sinusoidal function plus a constant. 
The resulting value of the amplitude of the sinusoid is $0.07 \pm 0.01$, whereas the period is $132.7 \pm 0.9$ days.

\section{Discussion and Conclusions}\label{sec:4}
We report the first measurement of the frequency derivative and the quasi-coherent timing solution of PSR J1023+0038 entirely obtained in the optical band.

As shown in the previous Section, the long gaps between observing runs prevented us from uniquely determining the differential number of phases between January and December 2018 ($N_1$), and February and November 2019 ($N_2$). This clearly means that the information on the continuous variation of the pulsar phase is lost. However, in the assumption that the pulsar has a characteristic phase and frequency noise not significantly larger than that measured in a single observing run (with the standard deviation of the phase measurements equal to $\sim$0.02, see Sect. \ref{sec:3.4}), fitting the available measurements with different combinations of $N_1$ and $N_2$ allows us to constrain the possible values that they can assume. Considering the four combinations of $N_1$ and $N_2$ with the lowest $\chi^2$ significantly reduces the indetermination on the timing solution, which is finally selected on the basis of a joint fit to all the available X-ray and optical frequency measurements. This timing solution (with $N_1=-2$ and $N_2=0$) shows the lowest phase and frequency residuals and, because of the way in which it is reconstructed, is referred to as `quasi-coherent'. It reproduces adequately well the evolution of the pulsar rotational phase and frequency measured in the optical band from January 2018 through January 2020, and gives a spin-down rate of $\dot \nu = (-2.53 \pm 0.04) \times 10^{-15}$ Hz$^2$.

We note that a certain level of phase noise is clearly present in the data of all five Aqueye+ runs. Fitting the phase residuals around the total timing solution with a sinusoidal function, we obtained that the phase noise has an amplitude of  $\sim$0.07. Timing irregularities of similar amplitude have been already observed in the radio band during the rotation-powered state. \citet{Archibald2013arx} showed that these irregularities have a periodic modulation caused by uncertainties in the knowledge of the orbital parameters of the system. Such kind of phase noise can in fact take place in our analysis, since we used a fixed epoch of the ascending node passage $T_{\rm asc}$ for each single run.

As it can be seen from Fig. \ref{fig:freq_aq_j16}, our best optical timing solution is in an agreement with past and current ``local'' measurements of the spin frequency made in the X-ray and optical bands, respectively. We found a value of $\dot \nu$ close to that measured in the radio band during the rotation-powered state $\dot \nu_{\rm r} = -2.3985 \times 10^{-15}$ Hz$^2$ (as reported in \citealt{Jaodand2016}). However, it is significantly smaller (by $\sim$20\%) than that obtained from the X-ray observations taken in 2013-2016 by \citet{Jaodand2016} ($\dot \nu_{\rm x} = (-3.0413 \pm 0.0090) \times 10^{-15}$ Hz$^2$). Although we cannot exclude the possibility that the value of $\dot\nu$ has changed with time, the currently available measurements of the frequency do not allow us to put a stringent constrain on $\ddot\nu$. Fitting jointly the X-ray and optical frequency measurements with a parabola function we obtained the best fitting $\ddot\nu = (2.5 \pm 1.0) \times 10^{-23}~{\rm Hz}^3$. We note that performing the fit of the optical phase measurements with a third-order polynomial function (introducing $\ddot\nu$), we obtained no timing solution which describes well both the optical phase measurements and the long-term evolution of the spin frequency. The X-ray timing solution extrapolated to the dates of the Aqueye+ runs is not consistent with the best optical timing solution as well.

The average spin rate of PSR J1023+0038 during the disc state is a crucial piece of information to understand the mechanism powering this source. The X-ray measurement from \citet{Jaodand2016} implies that the pulsar is spinning down at a rate 26.8\% faster than that measured during the radio pulsar phase. \citet{Haskell2017} proposed that this increase in spin-down rate is compatible with gravitational wave emission induced by asymmetries in pycno-nuclear reaction rates in the crust leading to a mass quadrupole large enough to account for the observed $\dot \nu$. 
We obtained that pulsar spin-down rate is only $\sim$5\% faster than $\dot \nu_{\rm r}$, which suggests that the losses due to emission of gravitational waves are probably lower than proposed in \citet{Haskell2017}.

Our results are more in line with the scenario reported in \citet{Papitto2019}. They show that optical and X-ray pulsations are likely to originate from a common underlying physical mechanism. They also propose that optical and X-ray pulses are synchrotron emission produced at the intrabinary shock that forms where a striped pulsar wind meets the accretion disk, within a few light cylinder radii away ($\sim$100 km), from the pulsar. In this scenario, the average pulsar spin down should be dominated by the magnetic dipole and pulsar wind emission, and hence be comparable to that measured during the radio pulsar phase.

Performing future observations with Aqueye+, it will be possible to continue monitoring an evolution of the pulsar spin-down. In addition, the larger data set will allow us to study the long-term orbital variations and, in particular, the time of the ascending node passage $T_{\rm asc}$ during the LMXB state.

\section*{Acknowledgements}
We thank the referee for his/her in-depth comments. This research is based on observations collected at the Copernicus telescope (Asiago, Italy) of the INAF - Osservatorio Astronomico di Padova. We acknowledge financial contribution from the ASI/INAF grant n. 2017-14-H.0 (projects ``High-Energy observations of Stellar-mass Compact Objects: from CVs to Ultraluminous X-Ray Sources'' and ``High-Energy observations of Stellar-mass Compact Objects: from CVs to the most luminous X-Ray Binaries''). A. P. and L. Z. also acknowledge financial support from grants ASI/INAF I/037/12/0 (PI: Belloni) and from INAF ``Sostegno alla ricerca scientifica main streams dell'INAF'', Presidential Decree 43/2018 (PI: Belloni). In this work we made use of the \texttt{XRONOS} software \citep{Stella1992} and of the following \texttt{Python} packages: \texttt{Matplotlib} \citep{Matplotlib2007}, \texttt{NumPy} \citep{NumPy2011}, \texttt{SciPy} \citep{SciPy2001} and \texttt{Astropy} \citep{Astropy2013}.

\section*{Data availability}
The data underlying this article will be shared on reasonable request to the corresponding author.

\bibliographystyle{mnras} 


\bsp	
\label{lastpage}
\end{document}